%% file: main.tex
\documentclass[]{spie}  
 
\usepackage{amsmath,amsfonts,amssymb}
\usepackage{graphicx}
\usepackage[colorlinks=true, allcolors=blue]{hyperref}

\usepackage{soul}
\usepackage{upgreek}
\usepackage{wrapfig}

\newcommand{\micron}{$\upmu$m}

\title{First operation of Transition-Edge Sensors in space with the Micro-X sounding rocket}

\author[a]{Joseph~S.~Adams}
\author[a]{Robert Baker}
\author[a]{Simon~R.~Bandler}
\author[b,h]{No\"emie~Bastidon}
\author[b,i]{Meredith~E.~Danowski}
\author[c]{William~B.~Doriese}
\author[d]{Megan~E.~Eckart}
\author[b]{Enectal\'i~Figueroa-Feliciano}
\author[b]{Joshua~Fuhrman}
\author[e,j]{David~C.~Goldfinger}
\author[e]{Sarah~N.T.~Heine}
\author[c]{Gene~C.~Hilton}
\author[d,*]{Antonia~J.F.~Hubbard}
\author[b]{Daniel~Jardin}
\author[a]{Richard~L.~Kelley}
\author[a]{Caroline~A.~Kilbourne}
\author[b]{Ren\'ee~E.~Manzagol-Harwood}
\author[f]{Dan~McCammon}
\author[a]{Takashi~Okajima}
\author[a]{Frederick~S.~Porter}
\author[c]{Carl~D.~Reintsema}
\author[a]{Peter~Serlemitsos}
\author[a]{Stephen~J.~Smith}
\author[g]{Patrick~Wikus}
\affil[a]{NASA Goddard Space Flight Center, 8800 Greenbelt Rd, Greenbelt, Maryland, USA}
\affil[b]{Northwestern University, 2145 Sheridan Road, Evanston, Illinois, USA}
\affil[c]{National Institute for Standards and Technology, 325 Broadway, Boulder, Colorado, USA}
\affil[d]{Lawrence Livermore National Laboratory, 7000 East Avenue, Livermore, California, USA}
\affil[e]{Massachusetts Institute of Technology, 77 Massachusetts Avenue,
Cambridge, Massachusetts, USA}
\affil[f]{University of Wisconsin, 1150 University Avenue, Madison, Wisconsin, USA}
\affil[g]{Bruker Biospin, Industriestrasse 26, 8117 Fällanden, Switzerland}
\affil[h]{Current affiliation: The Australian National University, Canberra ACT 0200, Australia}
\affil[i]{Current affiliation: Ball Aerospace, 1600 Commerce Street, Boulder, Colorado, USA}
\affil[j]{Current affiliation: Harvard and Smithsonian Center for Astrophysics, 60 Garden Street, Cambridge, Massachusetts, USA}

\authorinfo{Further author information: (Send correspondence to A.A.A.)\\A.A.A.: E-mail: aaa@tbk2.edu, Telephone: 1 505 123 1234\\  B.B.A.: E-mail: bba@cmp.com, Telephone: +33 (0)1 98 76 54 32}

\authorinfo{*Address correspondence to Antonia J.F.Hubbard: \href{hubbard8@llnl.gov}{hubbard8@llnl.gov} }

\begin{document} 
\maketitle

\begin{abstract}

With its first flight in 2018, Micro-X became the first program to fly Transition-Edge Sensors and their SQUID readouts in space. The science goal was a high-resolution, spatially resolved X-ray spectrum of the Cassiopeia~A Supernova Remnant. While a rocket pointing error led to no time on target, the data was used to demonstrate the flight performance of the instrument. The detectors observed X-rays from the on-board calibration source, but a susceptibility to external magnetic fields limited their livetime. Accounting for this, no change was observed in detector response between ground operation and flight operation. 
This paper provides an overview of the first flight performance and focuses on the upgrades made in preparation for reflight. The largest changes have been upgrading the SQUIDs to mitigate magnetic susceptibility, 
synchronizing the clocks on the digital electronics to minimize beat frequencies, and replacing the mounts between the cryostat and the rocket skin to improve mechanical integrity. As the first flight performance was consistent with performance on the ground, reaching the instrument goals in the laboratory is considered a strong predictor of future flight performance. 
\end{abstract}


\clearpage 
\input{sec_intro}
\input{sec_first_flight}
\input{sec_second_flight}
\input{sec_projections}


\bibliography{report} 
\bibliographystyle{spiebib} 

\acknowledgments
We gratefully acknowledge the technical support of Ernie Buchanan, John Bussan, Travis Coffroad, Sam Gabelt, Rob Hamersma, Kurt Jaehnig, Frank Lantz, Ken Simms, Tomomi Watanabe, George Winkert, and the WFF team. Micro-X operates under NASA Grant 80NSSC18K1445. Part of this work was performed under the auspices of the U.S. Department of Energy by Lawrence Livermore National Laboratory under Contract DE-AC52-07NA27344.

\end{document}

%% file: sec_intro.tex

\section{Introduction}
\label{sec:intro}
Micro-X is a sounding rocket mission designed to obtain high-resolution spectra of extended X-rays sources. It is also intended to further the technology readiness of Transition-Edge Sensor (TES) microcalorimeters and Superconducting QUantum Interference Devices (SQUIDs) for space operation. The instrument must fly on a sounding rocket because X-rays in the signal bandpass are attenuated by the atmosphere. Sounding rockets are suborbital missions, and each flight provides approximately five minutes of observation above an altitude of 160~km. Sounding rockets provide a good testbed for new technologies, as with Micro-X, because the instrument can be recovered after flight, providing the opportunity to modify the instrument for any future flights. Specialized engineering is required to successfully operate sensitive detectors, like microcalorimeters, in the vibrationally intense and time-constrained conditions of a sounding rocket flight \cite{Hubbard:2020, McCammon:2002}. 

This paper reviews the design (\S\ref{sec:instrument}) and performance (\S\ref{sec:first_flight}) of the instrument in the first flight, with particular emphasis on the impact of external magnetic fields (\S\ref{sec:first_flight_magnetic}) and instrument configuration (\S\ref{sec:first_flight_noise}) on performance. The instrument performance in flight and post-flight testing drove the modifications that have been implemented in preparation for the reflight (\S\ref{sec:second_flight}). A comprehensive description of the instrument and first flight performance are provided in \cite{Hubbard:2020}. 

%% file: sec_first_flight.tex
\section{First flight instrument}
 \label{sec:instrument}
  
\begin{figure} [htp]
	\begin{center}
		\includegraphics[width=\textwidth]{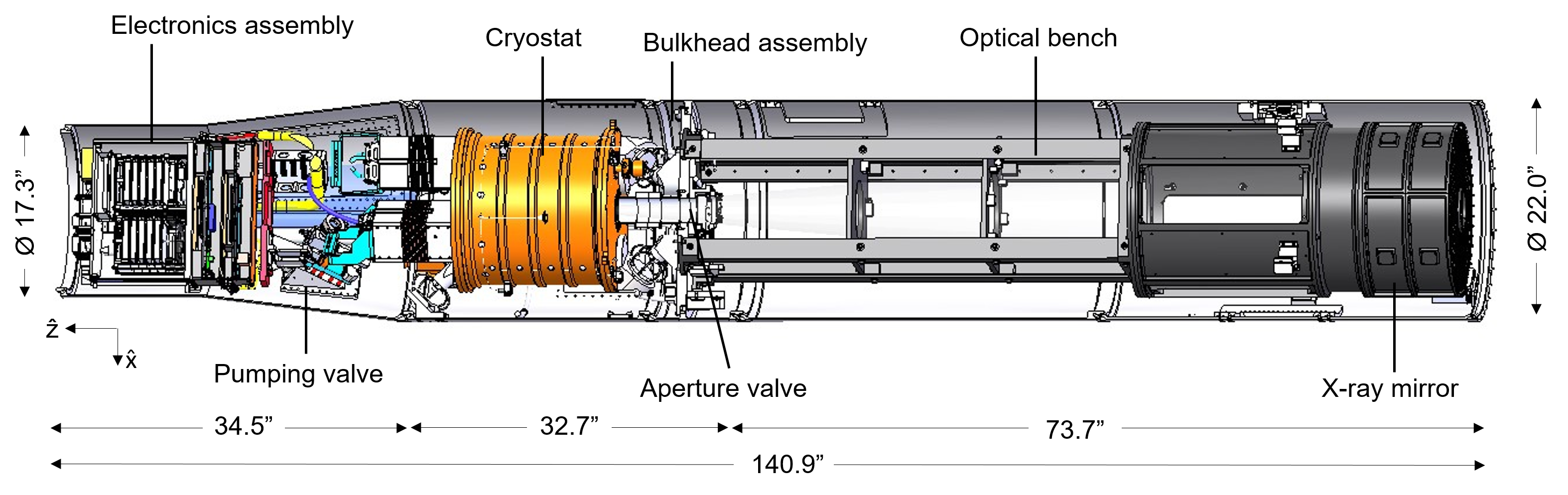}
	\end{center}
	\caption{X-rays enter the Micro-X instrument from the aft side (right in figure). They are focused by the X-ray mirror onto the detector array. The array sits inside the cryostat, with temperature control maintained by the pumped liquid helium ADR. On-board electronics control the state of the system and read out data.} 
	\label{fig:instrument}
\end{figure}

 The Micro-X instrument, shown in Figure~\ref{fig:instrument}, combines an X-ray mirror and a TES microcalorimeter array to achieve the instrument specifications provided in Table~\ref{table:instrument_specs}. Astronomical X-rays are focused by the X-ray mirror onto the detector array, which sits inside the cryostat. The detectors require a pumped liquid helium Adiabatic Demagnetization Refrigerator (ADR) to maintain a stable 75~mK~$\pm$~10~$\upmu$K operating temperature. The on-board electronics maintain the instrument operating conditions by executing flight operations and running the control loops for the ADR and SQUIDs. 

\begin{wraptable}{r}{0.5\textwidth}
   \vspace{-20pt}
    \begin{tabular}{ll}
    \hline
		\textbf{Parameter} & \textbf{Specification}\\
		\hline
		Array size & 7.2$\times$7.2~mm$^2$\\
		Effective area (1~keV) & 300\,cm$^2$\\
		Field of view & 11.8'\\
		Angular resolution (HPD) & 2.4 arcminutes\\
		Energy bandpass & 0.3 - 2.5~keV \\
		Quantum efficiency (0.2 - 4.0~keV) & $\sim$100\%\\
		Spectral resolution (0.2 - 3.0~keV) & 4.5~--~11~eV\\
	   	\hline
    \end{tabular}
	\caption{The Micro-X instrument specifications are optimized for high-resolution observations of extended X-ray sources. }
	\label{table:instrument_specs}
\end{wraptable}
The detector array is comprised of 128~TES microcalorimeter pixels \cite{Eckart:2009, Eckart:2013}. Microcalorimeters measure X-ray energy by tracking the temperature of an X-ray absorber. The absorber heats up when hit by an X-ray, then it cools down as heat dissipates through a weak thermal link into a cold bath. The rise and decay times of the thermal pulse are engineered when designing the pixel, and the pulse height (i.e., temperature rise) is proportional to the X-ray energy. 

\clearpage 
 TESs are thin, superconducting films that are biased into the transition region between their superconducting and normal states. In this regime, a small change in the temperature of the absorber leads to a large change in resistance of the TES, providing a high-resolution temperature measurement. Each Micro-X pixel uses a 140$\times$140~\micron$^2$ Mo/Au TES to track the temperature of a 590$\times$590~\micron$^2$ Bi/Au absorber. 

The TES array is read out with SQUIDs, which are low-noise, magnetically sensitive current amplifiers.  A three-stage DC SQUID Time-Division Multiplexing (TDM) readout scheme amplifies the TES signals to a level compatible with the room temperature electronics \cite{Chervenak:2000, deKorte:2003, Reintsema:2009}. The TES array is divided into two independent, 64-pixel TDM readout chains, each with four ``columns" of 16 ``rows." \cite{Reintsema:2009} The biases to the first-stage SQUIDs (SQ1) are cycled sequentially to turn on one row at a time. All 16~SQ1s in a column are connected in series to a summing coil, which couples the signal from the biased SQ1 to the second-stage SQUID (SQ2). The signal from the SQ2 is amplified by the third-stage Series Array (SA), which amplifies the signal by a factor of 100 before it reaches the room-temperature electronics. The room-temperature electronics bias each component, process the digital feedback loops for the SQUIDs, and record the science data. 

\begin{wrapfigure}{r}{0 pt}
		\includegraphics[width=0.55\textwidth]{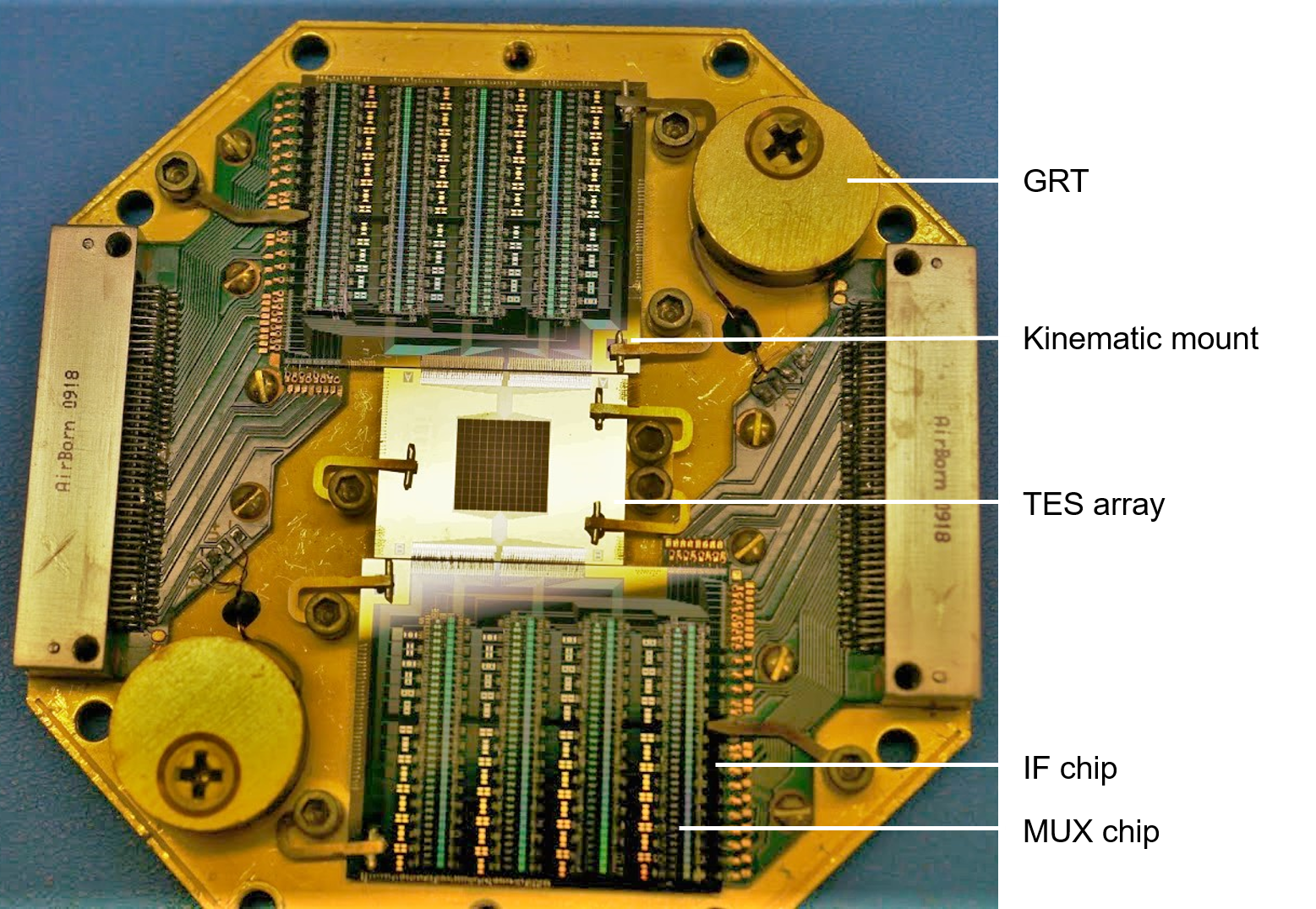}
	\caption{The FEA holds the TES array and the first two SQUID stages. The GRTs are used by the ADR to keep the FEA at a stable 75~mK. } 
	\label{fig:fea}
\end{wrapfigure}
The TES array and the first two SQUID stages (SQ1 and SQ2) are kinematically mounted to the Front End Assembly (FEA), shown in Figure \ref{fig:fea}. The TES array sits on a chip in the center of the FEA. The SQ1 and SQ2 are lithographically fabricated onto a set of multiplexer (MUX) chips. The MUX chips are glued to an interface (IF) chip, which holds bandwidth-limiting inductors and shunt resistors. There are two sets of MUX and IF chips, one for each science chain. The FEA base is a magnesium plate that is thermally connected to the ADR salt pill to maintain a stable operating temperature. Two Germanium Resistance Thermometers (GRTs) are used for temperature control and monitoring. 

The FEA must be shielded from infrared radiation, vibration, and magnetic fields. Six infrared blocking filters sit between the FEA and the cryostat aperture. They minimize the transmission of blackbody radiation from the warmer stages of the instrument while allowing 78\% of X-rays through \cite{Hubbard:2020}. Three stages of vibration isolation keep heat from coupling into the detector thermal stage during flight. These stages include: (1) wire rope isolators that mount the cryostat to the rocket (44~Hz resonant frequency); (2) G10 plastic tubes that support the liquid helium tank inside the cryostat (96~Hz resonant frequency); (3)
Kevlar suspension that hangs the detectors from the liquid helium tank (325~Hz resonant frequency) \cite{Danowski:2016}. The increasingly high frequencies of the inner stages dampen the outer stages for an effective isolation scheme. 

TESs and SQUIDs are sensitive to magnetic fields, and three stages of shielding minimize the effective field at the detector array, due to both the Earth's magnetic field and the ADR magnet. An external field coupling into the SQUIDs can move their response away from their stable operating point, and even outside the range of their flux-locked loops. The SQ2 stage is projected to have a larger magnetic susceptibility than the SQ1 stage due to the difference in their effective areas (468.5~\micron{}$^2$ and 5.0~\micron{}$^2$, respectively) \cite{Stiehl:2011}. The first stage of magnetic shielding is a bucking coil between the ADR magnet and the FEA. The second stage is a superconducting Nb can that surrounds the FEA to reject external magnetic fields through the Meissner-Ochsenfeld effect \cite{Meissner:1933}. In addition to the openings for X-ray entry and the thermally conductive rod to the ADR, the shield has two holes for harnessing to the FEA. The final stage of shielding is a field coil inside the Nb shield to counteract any trapped static field. It was not used in the first flight. The SA are not mounted on the FEA; they are mounted at the 2~K stage, and a separate Nb box surrounds them. In addition to these shielding layers, a removable Metglas blanket is wrapped around the cryostat each time it is cooled down from room temperature. This is done to minimize the amount of the Earth's magnetic field that freezes into the Nb shield when it goes superconducting.

To monitor detector response in flight, a radioactive calibration source inside the cryostat provides a continuous source of X-rays just outside the science band. It is mounted on the inside of the Nb can lid, roughly 1~cm from the detector array. The calibration source uses a ring of $^{55}$Fe to illuminate a KCl scintillator ring, producing Cl and K lines. The source rate in flight was measured to be $0.7$~counts/pixel/s, corresponding to a K-K$\upalpha$ rate of 0.23~$\pm$~0.06~counts/pixel/s. 

The rocket payload includes both the science instrument and several standard payload systems that are supplied by the NASA Sounding Rocket Operations Contract (NSROC) \cite{SRHB}. These systems include the celestial Attitude Control System (ACS) for pointing, the S-19 boost guidance system for trajectory control, the Ogive Recovery System Assembly (ORSA) to control the descent, and three telemetry systems for data transmission.  

\section{Flight performance}
 \label{sec:first_flight}

Micro-X launched from the White Sands Missile Range in New Mexico at midnight on July 22, 2018. The science target was the Cassiopeia~A Supernova Remnant. In flight, an ACS failure led the payload to spin and not spend any time on target. Despite this, the flight data was used to analyze the performance of the instrument. All flight operations were successfully executed, and the full data stream was recovered from on-board flash memory. The instrument achieved a stable 75~mK environment during the science observation, demonstrating the successful performance of both the ADR and the vibration isolation system. Despite the lack of a bright astrophysical target, the flight performance of the detectors was analyzed with X-rays from the calibration source (\S\ref{sec:first_flight_magnetic}--\ref{sec:first_flight_noise}). Details on the flight performance can be found in \cite{Hubbard:2020}. 

\begin{figure} [htp]
	\begin{center}
		\includegraphics[width=0.91\textwidth]{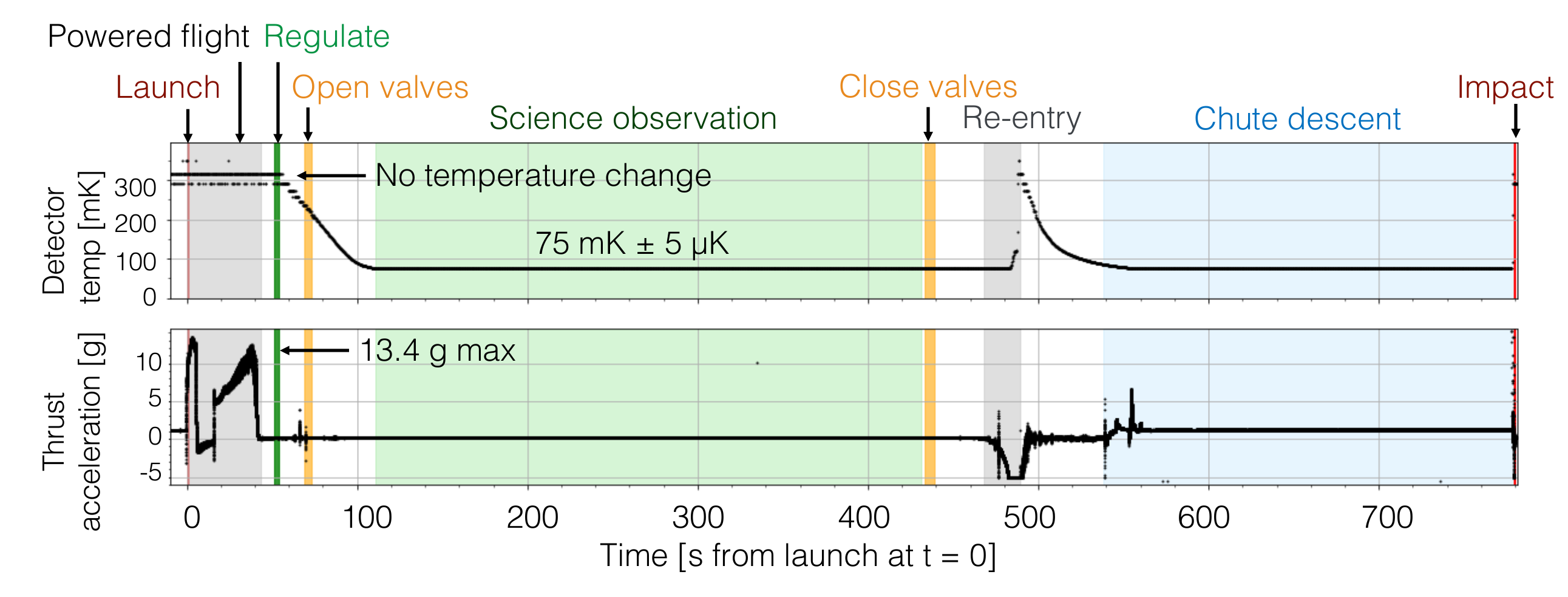}
	\end{center}
	\caption{The thermal performance of the instrument was successful. The detector stage was launched at 300~mK, and no discernible heat coupled into the stage during powered flight, which saw 13.4$g$ thrust acceleration on the exterior of the rocket. Stable temperature regulation of 75~mK$\pm$5~$\upmu$K was achieved during the science observation. } 
	\label{fig:flight_thermal}
\end{figure}

The rocket launched with the instrument in a protected configuration. The ADR was holding at 300~mK to maximize the available cooling power, and the instrument was closed off to space. The shutter door to the optics was closed, as was the aperture valve on the cryostat that exposes the detector array to the optics. Figure~\ref{fig:flight_thermal} shows that no discernible heat coupled into the detector stage (top) during powered flight (bottom). Once the payload was through powered flight, timer commands were sent to the instrument electronics to prepare for the science observation. The ADR began tight temperature control at 75~mK, and the apertures to the optics and the cryostat opened to expose the detector array to space. This was the start of the science observation. The observation lasted for 320~s, after which point the apertures closed in preparation for atmospheric re-entry. An apparent heat deposition during re-entry, likely from a glue failure in the wire rope isolators (\S\ref{sec:second_flight_additional_changes}), was successfully mitigated by the ADR, and tight temperature control was then maintained until impact. At 4.6~km above the ground, the parachute deployed for the remainder of the descent. 
 
\subsection{SQUID magnetic susceptibility}
\label{sec:first_flight_magnetic}

\begin{wrapfigure}{r}{0.5\textwidth}
   \vspace{-10pt}
		\includegraphics[width=0.5\textwidth]{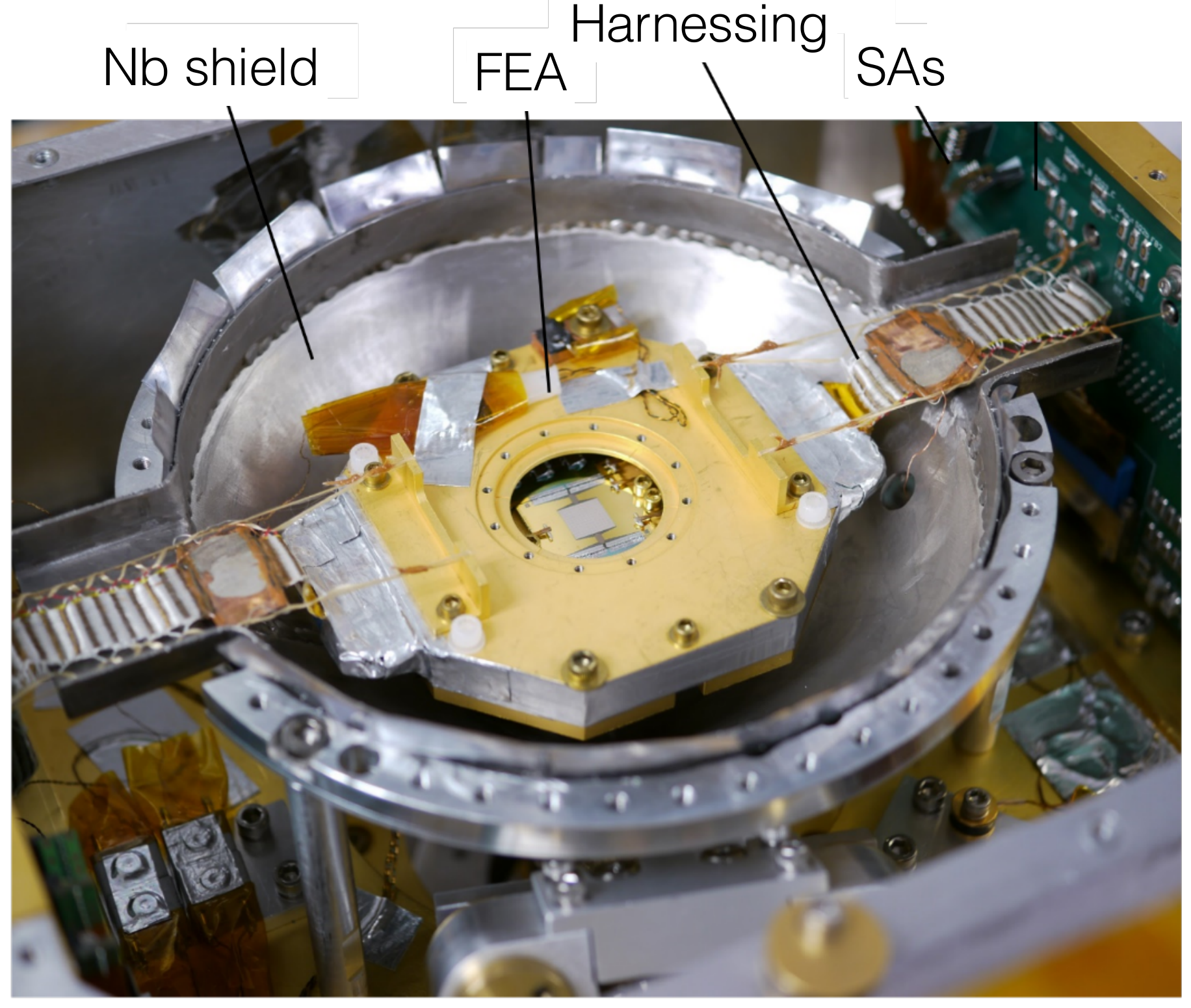}
	\caption{A superconducting Nb magnetic shield surrounds the FEA. The SQUIDs exhibited the largest magnetic susceptibility in the direction of the holes in the shielding for the harnessing. } 
	\label{fig:insert}
\end{wrapfigure}

The magnetic susceptibility of the SQUIDs to the Earth's field was illustrated using data from powered flight, when the TESs were normal. During this period, the rocket was spinning in the Earth's magnetic field, as measured by an on-board three-axis magnetometer. The SQUIDs response during this stage tracked the magnetometer reading, demonstrating a magnetic susceptibility, likely at the SQ2 stage. The SQUIDs were the most susceptible to the Earth’s field in the direction of the holes in the superconducting Nb shield for FEA harnessing, shown in Figure~\ref{fig:insert}.

The external magnetic field moved some SQUIDs far enough from their operating points that they unlocked, meaning that they effectively stopped reading out. This unlocking is qualitatively consistent with models simulating the response of the 3-stage SQUID readout to magnetic fields \cite{Hubbard:2020}. Under a normal flight trajectory, the payload would have been effectively stationary during the science observation. Due to the ACS failure, the payload continued to tumble throughout the science observation, so the SQUIDs continued to experience a changing field. As a result of this unlocking, the science observation had an exposure equivalent to 118 seconds of the full array. 

This magnetic susceptibility was confirmed with post-flight Helmholtz coil testing. The coils were placed along three orthogonal axes around the exterior of the rocket skin, with a magnetometer recording the applied field. They simulated spinning in the Earth's field by sweeping the magnetic field by one Earth field in either direction. The SAs were not discernibly susceptible to the external field, indicating successful magnetic shielding at the SAs. Testing of the entire 3-stage SQUID readout demonstrated that a large fraction of the SQUIDS reliably unlocked when exposed to simulated spinning. In addition to the field strength, a strong dependence on the speed of the magnetic field change (dB/dt) was observed. A gradual progression up to one Earth field moved the average composite SQUID response by 0.1$\Phi_0$, where $\Phi_0$ represents one period in the oscillatory SQUID response. By contrast, a discrete jump from no field to three Earth fields shifted the average composite SQUID response by 0.7$\Phi_0$. When the Helmholtz coils returned to zero field, the SQUID response did not return to its original position, but showed a significantly hysteretic response. This complicates comparisons to flight data. The susceptibility in the direction of the holes in the Nb shield was not probed directly; the closest measurement was clocked 22.5$^\circ$ from that direction. This was the most susceptible of the measured directions, consistent with the flight observation that the most susceptible direction was that of the Nb shield holes. 

\subsection{System noise}
\label{sec:first_flight_noise}

\begin{wrapfigure}{r}{0.45\textwidth}
   \vspace{-10pt}
		\includegraphics[width=0.45\textwidth]{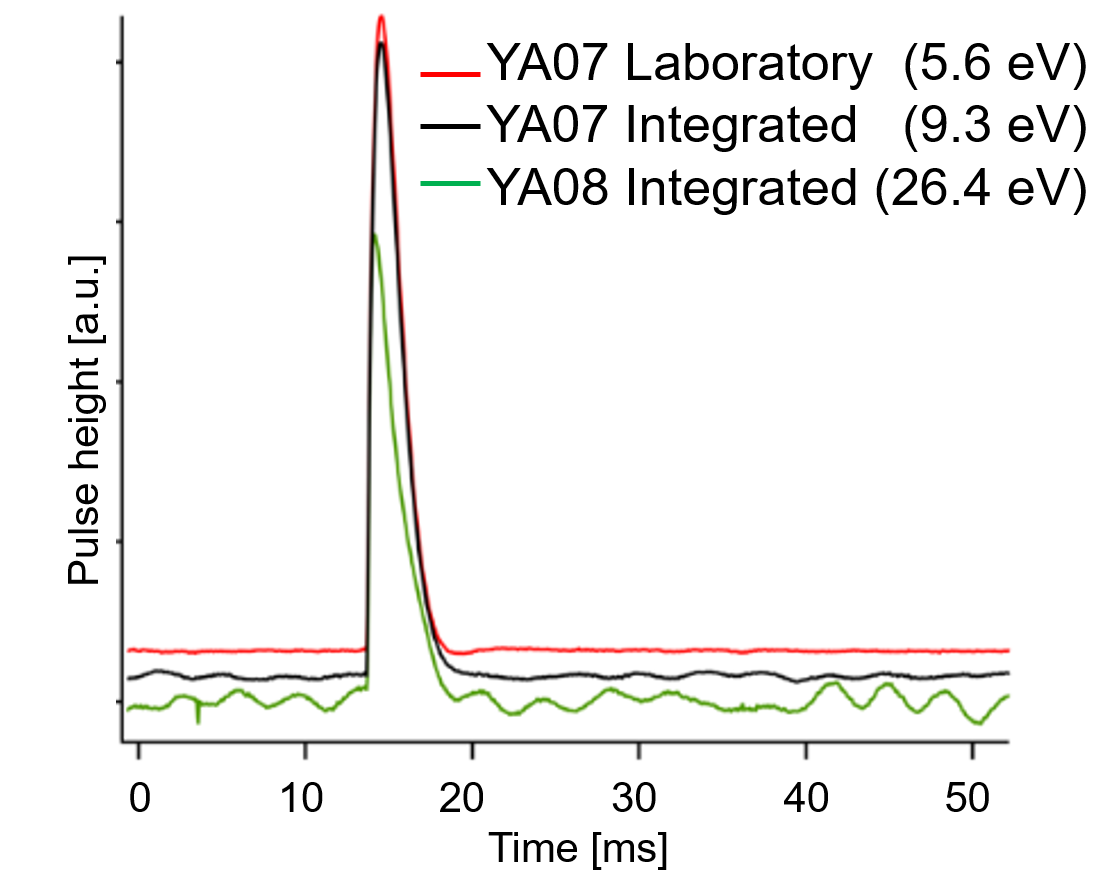}
	\caption{Pre-flight testing showed changes in detector performance, presumably driven from beating between the two science chain clocks, once the instrument was integrated. YA07 shows a moderate change in noise between the configurations, while YA08 shows significant noise. Post-flight testing confirmed that synchronizing the science chain clocks mitigates the observed beating. The pulses have been offset for clarity. } 
	\label{fig:nep_evolution}
\end{wrapfigure}

Integrating the instrument impacted detector performance, as described in detail in \cite{Hubbard:2020}. The fundamental detector resolution, in ideal laboratory conditions, was measured to be 4.5~eV. When running with one flight science chain, the average resolution of across the array was 6.5~eV, with 104 of the 116 active pixels achieving resolutions better than 10~eV. The instrument was then integrated, which means that the electronics were mounted above the cryostat and both science chains were turned on. In this configuration, the average resolution across the array increased to 11~eV, with 41~pixels achieving resolutions better than 10~eV. Figure~\ref{fig:nep_evolution} shows the difference in the two configurations for one pulse (YA07), which saw a moderate increase in system noise. A second pixel (YA08) picked up higher levels of system noise, with a corresponding degradation in resolution. A significant portion of the degradation is hypothesized to be from non-stationary noise due to beating between the clocks of the two science chains. Each science chain received its own clock from an MV encoder, and these clocks were not synchronized for the first flight. 

The flight performance is consistent with the ground performance in the integrated configuration, excluding the magnetic field effects described in \S\ref{sec:first_flight_magnetic}. No change to the noise spectrum or pulse shape was observed between the two datasets, as shown for a single pixel in Figure~\ref{fig:nep_evolution}. Additionally, no change in the noise spectrum or pulse shape was observed in flight between the array being open to space and the apertures to space being closed. This implies that performance improvements implemented on the ground are likely to predict flight performance.

Post-flight noise testing was consistent with the hypothesis that the science chain clocks were largely responsible for the observed beating. Laboratory testing used encoder simulators rather than the flight MV encoders, introducing uncertainty in the predicted flight response based on this testing. The beating observed in flight was qualitatively reproduced with the encoder simulators, and it was only observed when both simulators were on. It was mitigated by feeding a single clock into both simulators, effectively synchronizing them. The laboratory setup used for this synchronization test introduced additional noise to the system during the test, so no measurement was made of the impact of this synchronization on the total noise floor. The MV encoders have been synchronized for the reflight, as described in \S\ref{sec:second_flight_additional_changes}. A definitive statement on the impact of this synchronization on detector performance will be made once the payload is fully integrated for the reflight. 

%% file: sec_second_flight.tex
\section{Second flight system}
\label{sec:second_flight}
Modifications to the instrument were motivated by performance in flight and in post-flight testing. To address the magnetic susceptibility of the SQUIDs, a newer generation of MUX chips was installed on the FEA, and the material was changed for both the FEA kinematic mounts and the spacer in the superconducting Nb magnetic shield (\S\ref{sec:second_flight_fea_changes}). To address system noise, the clocks on both science chains were synchronized, and to avoid thermal failure, the wire rope isolators were replaced with a higher temperature version of the same model (\S\ref{sec:second_flight_additional_changes}). Each of these modifications has been implemented. 

\subsection{Magnetic Susceptibility Changes}
\label{sec:second_flight_fea_changes}

\begin{wrapfigure}{l}{0.38\textwidth}
		\includegraphics[width=0.38\textwidth]{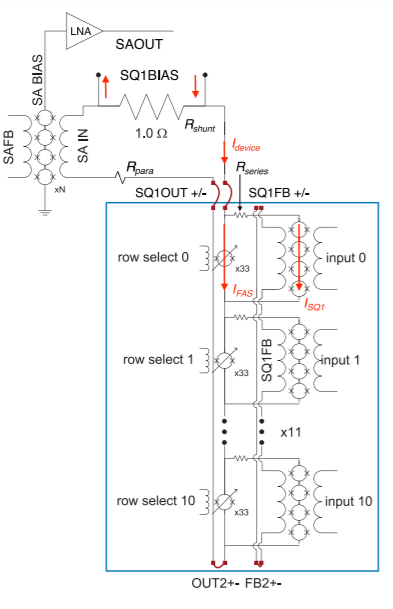}
	\caption{The MUX18b readout scheme uses flux-activated switches in the SQ1 stage and has demonstrated lower magnetic susceptibility than the MUX06a flown in the first Micro-X flight. Image from \cite{Reintsema:2019}.}
	\label{fig:new_mux}
\end{wrapfigure}

The MUX chips have been upgraded from NIST MUX06a to NIST MUX18b, which have a demonstrated lower magnetic susceptibilty \cite{Reintsema:2019}. The new readout scheme is shown in Figure~\ref{fig:new_mux}. It uses flux-activated SQUIDs in the SQ1 stage, rather than the current-activated SQ1s used in the MUX06a sheme. The TES array is the same as the first flight. The impact of the new SQUIDs on the noise floor is under investigation. 

In addition to the MUX chip replacement, the kinematic mounts that hold the TES and SQUID chips were replaced. The first flight kinematic mounts were made of tungsten carbide, which had become magnetized. The new mounts are made of sapphire. There is no indication that the old mounts impacted performance, but the modification was made to prevent any possible impact. 

The spacer on the inside of the superconducting Nb magnetic shield was replaced to improve the efficacy of the shield. The first flight spacer was made of stainless steel. It was replaced with Pb, which is superconducting at the instrument's operating temperature. The Nb is designed cover the entire spacer, but this replacement will provide a more robust magnetic shield if there are any gaps in the Nb layer. 

These three changes were implemented at the same time, and the combination of these changes significantly reduced the SQUID susceptibility to external magnetic fields. The Helmholtz coil testing that unlocked the first flight SQUIDs in the simulated Earth's magnetic field did not unlock the new SQUIDs. Introducing an external field equivalent to one Earth field moved the SQUID response by 0.016$\Phi_0$. The SQUIDs maintained a stable lock throughout testing. The SQUID response is independent of the rate of the change of the magnetic field, and no hysteresis is observed. The tests were performed with the TESs both on and off; no difference was observed between configurations. If the flight performance continues to follow the ground performance, these SQUIDs are not projected to unlock in flight. 

\subsection{Additional Changes}
\label{sec:second_flight_additional_changes}

The MV encoder clocks were synchronized to eliminate the beating noise that degraded detector performance in the integrated configuration (\S\ref{sec:first_flight_noise}). The telemetry engineers at the Wallops Flight Facility synchronized the encoders, and the new system passed handshake testing with the instrument electronics. Micro-X will be the first program to fly synchronized MV encoders. The impact of the synchronization on detector performance will be measured once the payload is integrated for the next flight. 

The wire rope isolators that mount the cryostat to the rocket skin experienced the only mechanical failure of the flight. The glue on the isolators is presumed to have failed due to its thermal contact with the rocket skin. This failure was reproduced in thermal testing in the laboratory after flight. The force required to compress the isolators by 1~mm was measured at 10$^\circ$C intervals from 20--170$^\circ$C, and the response was observed to degrade above 60$^\circ$C. In flight, the skin reached 60$^\circ$C during powered flight, and it stayed above this temperature through impact. The isolators were therefore replaced with a higher-temperature-rated version of the same model to maintain their effectiveness as part of the vibration isolation scheme. The new isolators passed thermal testing up to 170$^\circ$C. This is acceptable for flight, as the rocket skin reached a maximum temperature of 110$^\circ$C in the first flight. 

\begin{figure} [htp]
	\begin{center}
		\includegraphics[width=0.5\textwidth]{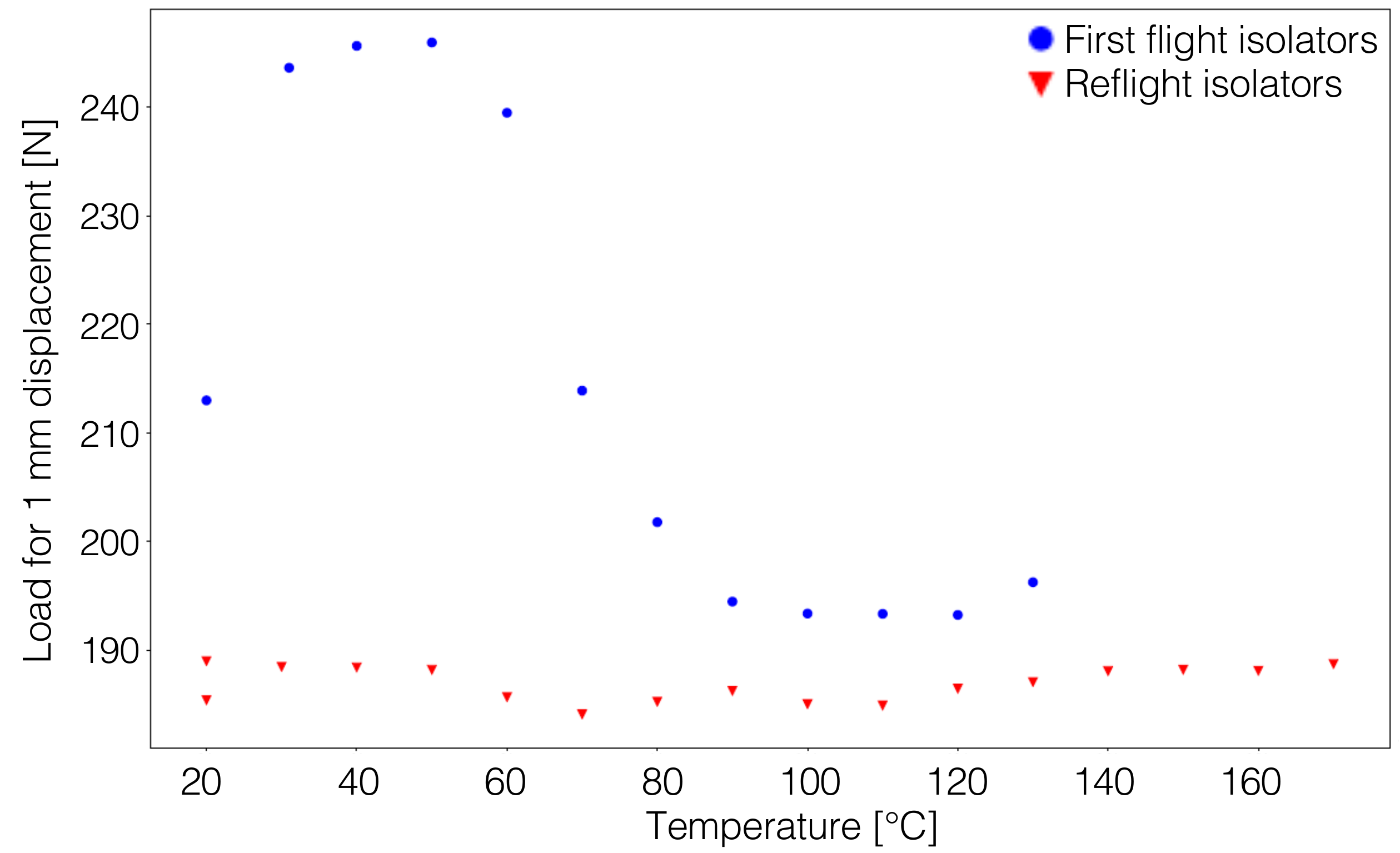}
	\end{center}
	\caption{The new wire rope isolators maintain a constant response up to 170$^\circ$C in laboratory testing. This is a significant improvement over the first flight isolators' response, which degraded above 60$^\circ$C. The rocket skin reached 110$^\circ$C in flight.  } 
	\label{fig:isolators}
\end{figure}

%% file: sec_projections.tex
\section{Conclusions}
\label{sec:future_projections}

The maiden flight of Micro-X saw the first operation of TESs and SQUIDs in space. Despite the ACS failure, the flight provided a measurement of the instrument flight performance. The flight operations and a number of instrument systems, notably the cryogenic system, had nominal performance. These systems are anticipated to fly again without any changes. The instrument flight performance was impacted by the magnetic susceptibility of the SQUIDs, increased system noise when the instrument was integrated, and the mechanical failure of the wire rope isolators. Each of these issues has been addressed by modifications in preparation for the next flight. 

The SQUIDs' magnetic susceptibility was addressed with a combination of new SQUIDs and material selection. The MUX chips were upgraded to NIST MUX18b. The kinematic mounts were converted from tungsten carbide to sapphire, and the spacer in the superconducting Nb magnetic shield was converted from stainless steel to lead. These changes were made at the same time, and the combination of them decreased the magnetic susceptibility to the point that the SQUIDs are not projected to unlock in future flights. The system noise was addressed by synchronized the encoder clocks that supply the clocks for the science chains. The wire rope isolator failure was addressed by replacing them with a higher-temperature-rated version of the same model. As the flight performance was consistent with the ground performance in the first flight, the ground performance following these upgrades is expected to predict future flight performance. Work is underway to characterize the noise performance of the system with the new SQUIDs, and the system will then move towards flight integration. 